\documentstyle[12pt]{article}
\newcommand{\absauthor}[1]{\small {\bf #1}}
\newcommand{\address}[1]{{\it #1}}
\newcommand{\be}{\begin{equation}}
\newcommand{\ee}{\end{equation}}
\newcommand{\me}{\medskip}
\newcommand{\no}{\noindent}
\newcommand{\pa}{\partial}
\newcommand{\ra}{\rightarrow}
\newcommand{\ci}[1]{\cite{#1}}

\begin{document}
\hspace{95mm} Phys. Rev. A {\bf 53}(1996)3798
\begin{center}
{\bf         Semiclassical wave equation and exactness of 
                         the WKB method } \\
\vspace{5mm}
\absauthor{              M. N. Sergeenko }  \\  
\vspace{2mm}
\address{The National Academy of Sciences of Belarus, 
           Institute of Physics, \\ Minsk 220072, Belarus }
\end{center}

\begin{abstract}
The exactness of the semiclassical method for three-dimensional
problems in quantum mechanics is analyzed. The wave equation
appropriate in the quasiclassical region is derived. It is shown that
application of the standard leading-order WKB quantization condition
to this equation reproduces exact energy eigenvalues for {\em all}
solvable spherically symmetric potentials.   \me

\no PACS number(s): 03.05.Ge, 03.65.Sq
\end{abstract}

\noindent {\bf 1. Introduction }\\

There are several problems in quantum mechanics that can be solved
exactly in terms of special functions. However, the same problems can
be solved exactly also in terms of elementary functions in the
framework of the WKB method.

The WKB method was originally proposed for obtaining approximate
eigenvalues of one-dimensional Schr\"odinger problems in the limiting
case of large quantum numbers. At present, the WKB method is
important and intriguing by its simplicity and efficiency, a powerful
tool of investigation not only in quantum mechanics but also in many
other branches of theoretical physics, for example, in the theory of
electromagnetic waves. In several common applications the method
gives very accurate results. However, from the moment of its
appearance up to now the same old problem of exactness of the WKB
approximation has arisen.

The structure of supersymmetric quantum mechanics motivates a
modified semiclassical quantization condition for one-dimensional
Hamiltonians \ci {1,Eck}. The supersymmetric WKB (SWKB) method is a
modification of the standard WKB quantization for obtaining the
quasiclassical eigenvalues of nonrelativistic Hamiltonians. It was
demonstrated \ci{Rep}-\ci{Kha} that the modified leading-order SWKB
quantization condition in each and every case reproduces the exact
energy eigenvalues for a class of solvable potentials. For these
models the solutions can be written in terms of elementary functions.

Recent successes of the SWKB quantization rule have revived interest
in the original WKB quantization condition. Proofs of varying degrees
of rigor have been advanced that demonstrate the exactness of the
standard WKB quantization condition \ci{Dunh}-\ci{Bru}. In Ref.
\ci{Bru}, a modification of the standard WKB approximation has been
considered for which the leading-order quantization condition
determines the exact energy eigenvalues for the same class of
solvable potentials. In this approach, exact eigenvalues have been
obtained by means of some phase distortions of WKB functions caused
by the potential singularities in the complex plane. The needed phase
distortions have been found with the use of quasiclassical
asymptotics of exact solutions.

The question of exactness of the WKB approximation is usually reduced
to the estimation of the high-order correction terms. The earliest
development of the WKB method for obtaining the high-order
corrections has been considered in Ref. \ci{Dunh}. Then, in
\ci{Krei}, the radial generalization of Dunham's one-dimensional WKB
quantization condition was derived with the help of the Langer
transformation \ci{Lang}. It was shown that the second- and
third-order integrals identically vanish for the hydrogen atom and
the three-dimensional harmonic oscillator \ci{Krei}.

One method (see, for example, Ref. \ci{Bru}) simply compares the
modified WKB result with the eigenvalues obtained from an exact
solution of the Schr\"{o}dinger equation. A second method (as in
\ci{Krei}) proceeds by showing that all additional high-order
correction terms to the WKB integral vanish for the given potential
\ci{Krei}. These proofs, however, are not entirely rigorous since
these correction terms are only asymptotically valid, i.e., as $\hbar
\ra 0$ \ci{Ros}. Furthermore, in the cases when a modified WKB
integral gives the exact eigenvalues, it is not even clear which
"correction" must be shown to be zero. Thus a different approach is
necessary if we are to prove the exactness of the leading-order WKB
quantization condition.

In this work we develop an approach to show the exactness of the
semiclassical approximation. We show that exact eigenvalues for the
class of solvable potentials can be reproduced by the {\em usual}
leading-order WKB quantization rule without any modification of the
method. Our approach to the problem under consideration differs
essentially from known ones \ci{Rep},\ci{Dunh}-\ci{Bru}, where the
one-dimensional problems have been considered.  Unlike the previous
approaches, instead of modification of the WKB method, we analyze an
original three-dimensional equation and begin our analysis from the
classic problem in the Hamilton-Jacobi formulation.

We analyze the semiclassical approximation in quantum mechanics using
two basic principles: the correspondence principle and the adiabatic
one. Starting from the three-dimensional classic problem and using
the two principles mentioned above, we derive  a wave equation
appropriate in the quasiclassical region. This "semiclassical" wave
equation has canonical form, i.e., it does not contain first
derivatives. Solving this equation for the spherically symmetric
potentials by the usual (to leading order in $\hbar $) WKB method, we
obtain exact energy eigenvalues for {\em all} spherically symmetric
potentials. The corresponding eigenfunctions have the same form as
the asymptotes of the exact solutions. \\

\noindent {\bf 2. WKB approximation for the radial equation}\\

Let us consider the Schr\"odinger equation for a spherically
symmetric potential $V(r)$,
\be (-i\hbar)^2\left[\frac{1}{r^2}\frac{\pa}{\pa r}
\left(r^2 \frac{\pa}{\pa r}\right)+\frac{1}{r^2\sin\,\theta}
\frac{\pa}{ \pa\theta}\left(\sin\,\theta\frac{\pa}{\pa\theta} \right)+
\frac{1}{r^2\sin^2\theta}\frac{\pa^2}{\pa\varphi^2} \right]\psi(\vec
r) =  2m[E-V(r)]\psi(\vec r).
\ee

If one substitutes $\psi (\vec r) = [U(r)/r]Y_{lm} (\theta ,\varphi)$
into Eq. (1), one obtains (after separation) the one-dimensional
problem for an effective potential $V_{eff}(r) =
V(r)+[l(l+1)\hbar^2/2mr^2]$:

\be \left[ \frac{d^2}{dr^2}+\frac{2m}{\hbar ^2}\left( E-V(r) -
\frac{l(l+1) \hbar ^2 }{2mr^2}\right) \right] U(r)=0.  \ee
\me \me

For the two-turning-point problems, the standard leading-order WKB
quantization condition is \ci{Shif}

\be
\int_{x_1}^{x_2}\sqrt{p^2(x)}dx=\pi\hbar\left(n+\frac 12\right),\ \ \
n=0,\,1,\,2,\,...,  \ee
where $x_1$,$x_2$ are the classical turning points,
$p^2(x)=2m[E-V(x)]$, and $V(x)$ is the potential. Application of the
quantization condition (3) to the radial equation (2) for the
solvable spherically symmetric potentials does not reproduce the
exact energy spectrum for the solvable potentials.

There is another difficulty arising in the semiclassical
consideration of the radial Schr\"odinger equation (2). This equation
has no the centrifugal term when $l=0$. This means that (i) the
effective potential $V_{eff}(r)=V(r)+[l(l+1)\hbar^2/2mr^2]$ has no
(for example, the Coulomb potential) minimum, (ii) the problem has no
left turning point and, as a result, (iii) we can not calculate the
ground state because we can not use the WKB quantization condition
(3) derived for two-turning-point problems.

In addition, the WKB solution of the radial Schr\"odinger equation is
irregular at $r\ra 0$, i.e., $R^{WKB}(r)\propto r^\lambda/\sqrt r$, 
$\lambda = \sqrt{l(l+1)}$, whereas the exact solution in
this limit is $R(r)\propto r^l$. In order for the first-order WKB
approximation to give the exact eigenvalues, the quantity $l(l+1)$ in
Eq. (2) must be replaced by $(l+\frac 12)^2$ \ci{Lang}.

The reason for this modification for the special case of the Coulomb
potential was pointed out in Ref. \ci{Lang} from the Langer
transformation

\be    r = e^x, \ \ \ \ \ U(r) = e^{x/2}X(x),    \ee
which mapped the point $r=0$ (for the radial problem) onto
$x=-\infty$ (for the one-dimensional one) and $r\ra \infty$ into $x
\ra \infty$ \ci{Krei}. As a result of such a transformation the wave
function $X(x)$ of the one-dimensional problem approaches zero for $x
\ra \pm \infty$, whereas the radial part of the solution $R(r)$
approaches zero for $r \ra 0$ and $\infty$. The effective potential
obtained when (4) is substituted into Eq. (2) takes the form

\be     V_{eff}(r)=V(r)+\frac{(l+\frac 12)^2\hbar^2}{2mr^2}.  \ee

The second- and third-order WKB corrections to the energy
quantization condition, derived in Ref. \ci{Krei}, have been shown to
be identically zero for the Coulomb potential. These corrections are
zero also for the harmonic oscillator. However, for other spherically
symmetric potentials, in order to obtain the appropriate Langer-like
correction terms, another special transformation of the wave function
and its arguments are required.

In the general case of spherically symmetric potentials, the
practical use shows that the quantization rule (3) for the effective
potential (5) yields in many cases exact energy eigenvalues. In
addition, the replacement $l(l+1)\ra (l+\frac 12)^2$ regularizes the
radial WKB wave function at the origin and ensures correct asymptotic
behavior at large quantum numbers.

In actual applications, the Langer correction in the radial
Schr\"odinger equation (2) is usually used without any proof
\ci{Kang}-\ci{Se2}.  Meanwhile, this correction has a deep physical
origin. In this work we give a foundation to the Langer replacement
and show the existence (in the quasiclassical region) of an integral
of motion, $\vec M^2=(l+\frac 12)^2\hbar ^2$. Our approach results in
an effective potential of the form (5), ensures the correct behavior
of the WKB wave function at $r\ra 0$, and provides its correct
asymptotic behavior. The quantization condition (3) gives the exact
energy spectrum.

In Ref. \ci{Bru}, exact eigenvalues for several potentials have been
obtained with the help of some phase distortions of WKB functions
caused by the potential singularities in the complex plane. Herein,
in Ref. \ci {Bru}, as in the case of Langer transformation, the
needed phase distortion is the "quarter".  In fact, this additional
constant $\frac 14$ changes (as will be shown below) the constant of
motion $\vec L^2=l(l+1)\hbar^2$ by $\vec M^2=(l+\frac 12)^2\hbar^2$.
Therefore, the phase distortion and Langer correction require the
same modification of the WKB solution, namely, the changing squared
angular momentum eigenvalues.  

The WKB solution of the angular Schr\"odinger equation has, analogous 
to the radial solution, incorrect behavior at $\theta\ra 0$:  
$\Theta^{WKB}(\theta)\propto\theta^\mu$, $\mu^2=m^2-\hbar^2/4$, 
while the exact regular solution in this limit is $\Theta_l^m(\theta )$ 
$\propto\theta^{|m|}$. Therefore the  angular equation also should be
modified in the quasiclassical region to have for the function
$\Theta^{WKB}(\theta)$ the same behavior at $\theta\ra 0$ as the
exact one.

In this paper we show that the main problem concerning the exactness
of the WKB approximation is to reduce the Schr\"odinger equation to
the "correct" canonical form, i.e., to the equation without first
derivatives.  In the case of the spherical coordinates, the Langer
transformation changes the centrifugal term $l(l+1)\hbar^2/r^2$ in
the Schr\"odinger equation by $(l+\frac 12)^2\hbar^2/r^2$. In fact, 
this requires changing the equation of motion in the quasiclassical
region. The required form of the wave equation and, as a result, the
centrifugal term can be obtained within the framework of the same
semiclassical approach. Below we deduce the so-called semiclassical
wave equation that has the necessary canonical form. \\

\noindent {\bf 3. Semiclassical wave equation}\\

One of the fundamental principles of quantum mechanics is the
correspondence principle, which has been used at the stage of the
creation of the quantum theory. The WKB method is the mathematical
realization of the correspondence principle and it is usually used
as a tool to obtain the approximate solution of the one-dimensional
Schr\"odinger equation in the quasiclassical region at large values
of quantum numbers.  However, this same principle is used to derive
the wave equation in quantum mechanics.

Consider the classical problem in the Hamilton-Jacobi formulation.
The static Hamil- ton-Jacobi equation for a particle of mass $m$
moving in the field of the spherically symmetric  potential $V(r)$
is

\be \left(\frac{\pa S_0}{\pa r}\right)^2+\frac 1{r^2}\left(\frac{
\pa S_0}{\pa \theta }\right)^2+\frac 1{r^2\sin^2\theta}\left(
\frac{\pa S_0}{\pa\varphi}\right)^2=2m\left[E-V(r)\right],  \ee
where $S_0$ is the classical action of the system.  The wave equation
in quantum mechanics can be obtained with the use of the classical
equation according to the correspondence principle:  $f\ra \hat f$,
where $f$ is the physical quantity and $\hat f$ is the corresponding
operator.

Let us write the equation corresponding to (6) in a quantum
mechanical form as

\be \left( \hat p_r^2+\frac{\hat p_\theta ^2}{r^2}+\frac{\hat
p_\varphi ^2}{ r^2\sin^2\theta }\right) \tilde \psi (\vec r)=\left[
E-V(r)\right] \tilde \psi (\vec r),   \ee
where $\hat p_q$ is the operator of the momentum conjugated with the
coordinate $q$ ($q=r,\,\theta ,\,\varphi$ for the spherical
coordinates). To find the apparent form of the operators $\hat p_q^2$,
let us represent the function $\tilde\psi(\vec r)$ in the form

\be
\tilde \psi (\vec r)=A\,\exp\left[\frac i\hbar S(\vec r)\right], \ee
where $S(\vec r)$ is the action in quantum mechanics and $A$ is the 
arbitrary constant. Consider the second derivatives of the function 
$\tilde\psi(\vec r)$. For the second derivative 
$\pa^2\tilde\psi/\pa r^2$ we have:

$$ \frac{\pa^2\tilde\psi}{\pa r^2}=\left[\left(\frac i\hbar
\frac{\pa S}{\pa r}\right)^2+\frac i\hbar\frac{\pa^2S}{
\pa r^2}\right]\tilde\psi  $$
or

\be \left(-i\hbar\frac\pa{\pa r}\right)^2\tilde \psi = \left[
\left(\frac{\pa S}{\pa r}\right)^2 + \frac\hbar i\frac{\pa ^2S}{\pa
r^2}\right]\tilde\psi . 
\ee

Now, according to the WKB method, substitute into (9) the
expansion of the action $S(\vec r)$ in powers of $\hbar $, $S(\vec
r)=S_0(\vec r)+$ $\hbar S_1(\vec r)+\hbar ^2S_2(\vec r)+...$ Then
we obtain to leading order in $\hbar $,

\be  \left(-i\hbar \frac \pa {\pa r}\right)^2\tilde \psi \simeq
\left( \frac{\pa S_0}{\pa r}\right) ^2\tilde \psi . 
\ee
Equation (10) is appropriate in the quasiclassical region where the
following condition is satisfied:

\be  \qquad \left( \frac{\pa S_0}{\pa r}\right) ^2\gg \hbar \left|
\frac{\pa ^2S_0}{\pa r^2}\right| .
\ee
By definition, the value $p_r=\pa S_0/\pa r$ is the radial
momentum.  Therefore the corresponding operator on the left-hand side
of Eq. (10) is the squared radial momentum operator, i.e.,

\be  \hat p_r^2 = \left(-i\hbar\frac{\pa}{\pa r}\right)^2.  
\ee

The form of the operators $\hat p_{\theta}^2$, $\hat p_{\varphi}^2$ is
found analogously:

\be  \hat p_{\theta}^2 = \left(-i\hbar\frac{\pa}{\pa\theta}
\right)^2,\ \ \ \ \ \hat p_{\varphi}^2 = \left(-i\hbar\frac
{\pa}{\pa\varphi}\right)^2,   \ee
and Eq. (7) takes the form:

\be  (-i\hbar )^2\left[\frac{\pa ^2}{\pa r^2}+\frac
1{r^2}\frac{ \pa ^2}{\pa \theta ^2}+\frac
1{r^2\sin^2\theta }\frac{\pa ^2}{ \pa \varphi
^2}\right] \tilde \psi (\vec r)=2m\left[ E-V(r)\right]
\tilde \psi (\vec r). \ee
Equation (14) is the second-order differential equation in canonical
form. This semiclassical wave equation is closely related to the
classical one (6) and is appropriate to describe quantum systems
in the quasiclassical region. It is easy to see that Eq. (14) can be
obtained from the classical Hamilton-Jacobi equation (6) with the
help of the correspondence principle

\be \frac{\pa S_0}{\pa q}\ra -i\hbar \frac \pa {\pa
\,q}, \ \ \ \ \ \ \ q=r,\,\theta ,\,\varphi . \ee
As will be demonstrated below, Eq. (14) is solved by the WKB method
in the elementary functions and the corresponding energy eigenvalues
coincide with the exact ones for the many solvable potentials.  The
correlation of the function $\tilde{\psi}(\vec r)$ with the wave
function ${\psi}(\vec r)$ in the Schr\"{o}dinger Eq. (1) is given by

\be
\tilde{\psi}(\vec r)=\sqrt{det\,g_{ij}}\psi(\vec r),  \ee
which follows from the identity: 
$\int\left|\psi(\vec r)\right|^2d^3\vec r$ $\equiv\int\left|\psi(\vec r)
\right|^2det\,g_{ij}
dr\,d\theta\,d\varphi $ $= 1$, where $g_{ij}$ is the metric tensor
(det$\,g_{ij}=r^2\sin\,\theta$ for the spherical coordinates). \\

\noindent {\bf 4. Semiclassical quantization in examples}\\

As is well known, working within the framework of the usual WKB
approximation, one does not get the exact spectrum from the
leading-order term for many solvable potentials such as Hulth\'en and
Rosen-Morse potentials unless one supplements it with Langer-like
correction terms that are different for different potentials.
However, application of the leading-order WKB quantization rule (3)
to the semiclassical Eq. (14) gives, as one can easily check, the
exact energy eigenvalues for all solvable spherically symmetric
potentials and no further Langer-like corrections are necessary.

To illustrate Eq. (14) consider several classic problems. For the
spherically symmetric potentials Eq. (14) is separated, yielding
three second-order equations

\be \left(-i\hbar\frac{d}{dr}\right)^2\tilde R(r) = \left[2m
\left(E-V(r)\right)- \frac{\vec M^2}{r^2}\right]\tilde R(r), \ee

\be \left(-i\hbar\frac{d}{d\theta}\right)^2\tilde{\Theta}(\theta) =
\left[ \vec M^2-\frac{M_z^2}{\sin^2\theta}\right] \tilde{\Theta}
(\theta), \ee

\be  \left(-i\hbar\frac d{d\varphi}\right)^2\tilde\Phi(\varphi
)=M_z^2\tilde\Phi(\varphi),
\ee
where $\vec M^2$, $\vec M_z^2$ are the constants of separation and,
at the same time, integrals of motion.  The squared angular momentum
$\vec L^2$ defined from the Schr\"odinger equation, takes the values
$\vec L^2=l(l+1)$ $\hbar^2$. In this section we show that, in the
quasiclassical region, the squared angular momentum takes other
values, namely $\vec M^2=(l+\frac 12)^2\hbar^2$. \newpage

\centerline{\bf  A. Angular momentum in the quasiclassical region}
\medskip

The WKB quantization rule has proven to find approximate eigenvalues
for the one-dimensional or radial Schr\"odinger equation. As for the
angular Schr\"odinger equation, the WKB quantization rule does not
reproduce exact eigenvalues. Besides, as mentioned above, the WKB
solution of the angular Schr\"odinger equation, $\Theta^{WKB}
(\theta)$, has incorrect behavior at $\theta \ra 0$ and $\pi$.

Let us deal with the angular Eqs. (18) and (19), which determine the
angular momentum and its projection in the quasiclassical region.
Equation (19) determines eigenvalues of the angular momentum
projection operator $\hat M_z^2=[-i\hbar (d/d\varphi)]^2$. The
solution of this equation, $\tilde\Phi_m(\varphi)$, is well known:
$\tilde\Phi_m(\varphi)$ $=\frac 1{\sqrt{2\pi}}\,e^{im\varphi}$,
$M_z=m\hbar$, $m=0,\pm 1,\pm 2,...$.

Equation (18) is especially important in our semiclassical approach
since it determines the squared angular momentum eigenvalues $\vec
M^2$ that enter into the radial equation (17). The WKB quantization
condition (3) appropriate to Eq. (18) and the WKB solution at the
interval $[\theta_1, \theta_2]$ are

\be \int_{\theta_1}^{\theta_2}\sqrt{p^2(\theta)}d\theta =
\pi\hbar\left(n_\theta + \frac 12\right), \ \ \ n_\theta = 0,1,2,..., 
\ee

\be 
\tilde{\Theta}^{WKB}(\theta) = \frac A{\sqrt{\left|p(\theta)\right|}}
\cos\left(\int_{\theta_1}^{\theta}\sqrt{p^2(\theta)}d\theta -
\frac\pi 4\right),  
\ee
where $p^2(\theta)=\vec M^2 - M_z^2/\sin^2\theta$; $\theta_1$,
$\theta_2$ are the roots (classical turning points) of the equation
$p^2(\theta)=0$ and $A$ is the arbitrary constant. Introducing a
variable $\alpha = \theta -\frac\pi 2$ the phase integral (20) can
be written in closed form

$$ \sqrt{\vec M^2-M_z^2} \int_{\alpha_1}^{\alpha_2}
\sqrt{1-k^2\sin^2\alpha} \frac{d\alpha}{\cos\,\alpha} = \pi
\left(\sqrt{\vec M^2} - \sqrt{M_z^2}\right),      $$
where $k^2 = \vec M^2/(\vec M^2-M_z^2)$. Setting $M_z = m\hbar$, we get,
for the $\vec M^2$,

\be \vec M^2=\left(l+\frac 12\right)^2\hbar^2, \ \ \ 
l=\left| m\right| + n_\theta.
\ee

Equation (22) represents the squared angular momentum eigenvalues in
the quasiclas- sical region. Since the eigenvalues (22) have been
obtained from the solution of the angular equation (18), this result
is appropriate for any spherically symmetric potential. As noted
above, the WKB solution $\Theta^{WKB}(\theta )$ of the angular
Schr\"odinger equation has incorrect asymptotes at $\theta\ra 0$ and
$\pi $. At the same time, the WKB solution (21) corresponding to the
eigenvalues (22) has the correct asymptotic behavior at these points
for all values of $l$. So far, as the momentum 
$p(\theta )\simeq \left| m\right|/\theta$ at $\theta \ra 0$, this gives, 
for the WKB solution in the representation of the wave function 
$\psi(\vec r)$ (see Eq. (16)), 
$\Theta_l^m(\theta)=\tilde\Theta^{WKB}(\theta)/
\sqrt{\sin\,\theta} \propto \theta ^{|m|}$ which corresponds to the
behavior of the exact wave function $Y_{lm}(\theta ,\varphi )$ at
$\theta \ra 0$.  The normalized quasiclassical solution (21) far from
the turning points, where $p(\theta)\simeq (l+\frac 12)\hbar $, has
the form

\be \tilde{\Theta}_l^m(\theta) = \sqrt{\frac 2\pi\frac{l+\frac 12}
{l-\left| m\right| + \frac 12}}\cos\left[\left(l+\frac 12\right)
\theta -\frac{\pi}2\left| m\right| -\frac{\pi}4 \right],
\ee
which, for the function $\Theta_l^m(\theta)$, agrees with the
asymptote of the spherical functions $Y_{lm}(\theta,\varphi)$.

Now, using the obtained solution of the angular semiclassical Eq.
(18), let us consider the radial Eq. (17). Substituting (22) into Eq.
(17), we obtain the radial semiclassical equation for the effective
potential (5) and no further Langer-like correction is necessary. The
leading-order WKB quantization condition (3) appropriate to the
radial Eq. (17) is

\be  \int_{r_1}^{r_2}\sqrt{p^2(r)}dr = 
\pi\hbar\left(n_r+\frac 12\right), \ \ \ n_r = 0,1,2,..., 
\ee
where the classical turning points $r_1$,$r_2$ are roots of the
equation

\be p^2(r)\equiv 2m[E-V(r)]-\frac{(l+\frac 12)^2\hbar ^2}{r^2}=0.
\ee

It is easy to check that the quantization condition (24) yields exact
energy eigenvalues for all solvable spherically symmetric potentials
$V(r)$ \ci{Kre,Ros}, such as the Coulomb potential, the
three-dimensional harmonic oscillator, and other ones. To demonstrate
Eq. (14), let us consider several potentials of interest.  \\

\centerline{\bf B. The Morse potential, $V(r)=V_0[e^{-2\alpha
(r/r_0-1)}-2$ $e^{-\alpha (r/r_0-1)}]$} \medskip

For this potential, let us consider, first, the radial Schr\"odinger
equation (2), which does not contain the centrifugal term at $l=0$:

\be \left(-i\hbar\frac d{dr}\right)^2U(r)=2m\left[E-V_0 e^{-2\alpha
(r-r_0)/r_0}+2V_0e^{-\alpha (r-r_0)/r_0}\right]U(r) = 0.  \ee

The first-order WKB quantization condition appropriate to this
equation is

\be \int_{r_1}^{r_2}\sqrt{2m[E-V_0e^{-2\alpha (r-r_0)/r_0} +
2V_0e^{-\alpha (r-r_0)/r_0}]}dr=\pi \hbar (n_r+\frac 12).  \ee
Introducing a variable $x=e^{-\alpha (r-r_0)/r_0}$, we reduce the
phase-space integral to the well known one. The sequential simple
calculations result in the exact energy eigenvalues

\be E_n = -V_0\left[1-\frac{\alpha\hbar(n_r+\frac
12)}{r_0\sqrt{2mV_0}} \right]^2.  \ee

Now, let us deal with {\em the semiclassical} equation (17) for this
potential, which (unlike the Schr\"odinger equation) contains the
non-vanishing centrifugal term $[\hbar^2/4r^2]$ at $l=0$:

\be \left( -i\hbar \frac d{dr}\right) ^2\tilde R(r)=\left[ 2m\left(
E-V_0e^{-2\alpha (r-r_0)/r_0}+2V_0e^{-\alpha (r-r_0)/r_0}\right) -
\frac{ (l+\frac 12)^2\hbar ^2}{r^2}\right] \tilde R(r)=0.  \ee
The WKB quantization condition (24) appropriate to Eq. (29) is:

\be I = \int_{r_1}^{r_2}\sqrt{2m\left( E-V_0e^{-2\alpha
(r-r_0)/r_0}+2V_0e^{-\alpha (r-r_0)/r_0}\right) -\frac{\lambda
^2}{r^2}} dr=\pi (n^{\prime }+\frac 12), \ee
where $\lambda =\hbar (l+\frac 12)$. In the region $r>0$, the problem
under consideration has two turning points $r_1$,$r_2$ which are
defined by Eq. (25).  To calculate this integral introduce the
variable $\rho =\frac r{r_0}$ and replace the integral along the
interval $[\rho_1,\rho_2]$ by the contour integral in the complex
plane $\rho $ with $Re\,\rho = r/r_0$, where the integral is taken
about a contour $C$ enclosing the classical turning points $\rho
_1$,$\rho _2$ and there are no other singularities of $p(r)$. Now,
using the method of stereographic projection, we should exclude the
singularities outside the contour $C$, i.e., at $\rho =0$ and 
$\infty$. Excluding these infinities we have, for the integral (30),

\be I=\frac 12\oint \sqrt{2mr_0^2[E-V_0e^{-2\alpha (\rho
-1)}+2V_0e^{-\alpha (\rho -1)}]-\frac{\lambda ^2}{\rho ^2}}d\rho
=\frac 12(I_1+I_2), \ee
where
$$
I_1=r_0\sqrt{2m}\oint_{C_1}\sqrt{E-V_0e^{-2\alpha (\rho -1)}+2V_0e^{-\alpha
(\rho -1)}}d\rho   $$
is reduced to the integral considered above and $I_2=\oint_{C_2}$
$\sqrt{-\lambda^2/\rho^2}d\rho = -2\pi \lambda $.

Therefore for the phase-space integral (31) we have
\be
I = -\pi\lambda -\frac{\pi r_0}{\alpha}\left(\sqrt{-2mE}-
\sqrt{2mV_0}\right) \ee
and for the energy eigenvalues this gives

\be E_n=-V_0\left[ 1-\alpha \frac{\hbar (n_r+\frac 12)+\lambda
}{r_0\sqrt{2mV_0}} \right] ^2.  \ee
Setting in (33) $\lambda =0$, we arrive at the formula (28) obtained
from the Schr\"odinger equation at $l=0$. However, in our case
$\lambda_{min}=\hbar /2$ at $l=0$ and the energy eigenvalues
are:

\be E_n=-V_0\left[1-\frac{\alpha\hbar
(n_r+1)}{r_0\sqrt{2mV_0}}\right]^2.  \ee

Formula (34) for $E_n$  is different from the expression (28)
obtained from the Schr\"odinger equation for the Morse potential at
$l=0$.  This difference is caused by the centrifugal term
$[\hbar^2/4r^2]$ in the radial semiclassical Eq. (17) at $l=0$.  Thus
we obtain two results for the Morse potential by the WKB method: the
known exact eigenvalues (28) obtained from the Schr\"odinger equation
and another result (34) obtained from solution of Eq. (17). \medskip

\centerline{\bf C. The Hulth\'en potential, $V(r)=-V_0e^{-r/r_0}/
(1-e^{-r/r_0})$}  \medskip

The Hulth\'en potential is known as nonsolvable by the standard WKB
method potentials, unless one supplements it with Langer-like
corrections.  However, solving the {\em semiclassical equation} (14)
for this potential by the usual WKB method, we obtain the exact
analytic result.

The leading-order quantization condition (24) for the Hulth\'{e}n
potential is

\be I=\int_{r_1}^{r_2}\sqrt{2m\left( E + V_0\frac{e^{-r/r_0}}
{1-e^{-r/r_0}}\right) -\frac{(l+\frac 12)^2\hbar ^2}{r^2}}dr=\pi
\hbar (n^{\prime }+\frac 12).  \ee
In the region $r>0$, this problem has two turning points $r_1$,$r_2$.
The phase-space integral (35) is calculated analogously to the above
case. Replace the integral along the interval $[r_1,\,r_2]$ by the
contour integral in the complex plane of the variable $\rho $, $\rho
=\frac r{r_0}$, where contour $C$ encloses the classical turning
points $\rho_1$, $\rho_2$.  Using the method of stereographic
projection, we should exclude the infinities outside the contour $C$.
Excluding these infinities we have, for (35),

\be I=\frac 12\oint \sqrt{2mr_0^2\left( E+V_0\frac{e^{-\rho
}}{1-e^{-\rho }}\right) -\frac{(l+\frac 12)^2\hbar ^2}{\rho ^2}}d\rho
=\frac 12(I_1+I_2), \ee
where

$$  I_1=\oint_{C_1}\sqrt{2mr_0^2\left( E+V_0\frac{e^{-\rho }}
{1-e^{-\rho }}\right) }d\rho ,   $$
and $I_2=i\hbar(l+\frac 12)\oint_{C_2}\frac{d\rho}\rho$.

To calculate the integral $I_1$, let us introduce the variable
$z=e^\rho -1$. Then the simple integration gives, for $I_1$,

\be
I_1=r_0\sqrt{2m}\oint_{C_2}\sqrt{E+\frac{V_0}z}\frac{dz}{z+1}\equiv
\ee
~~~~~~~~~~~~~~~
$$
r_0\sqrt{2m}\left( \oint_{C_\infty }\sqrt{E+\frac{V_0}z}dz -
\oint_{C_{-1}}\sqrt{Ez^2+V_0z}\frac{dz}{z+1}\right) =      $$
~~~~~~~~~~~~~~~
$$ 2\pi r_0\sqrt{-2m}\left[ -\sqrt{-E}+\sqrt{-E+V_0}\right].  $$
Substituting the integration result into Eq. (35), we immediately get the
exact energy spectrum

\be E_n=-\frac 1{8mr_0^2}\left(\frac{2mV_0r_0^2}N-N\right)^2, 
\ee
where $N=(n^{\prime}+l+1)\hbar $ denotes the principal quantum number.

Thus application of the standard leading-order WKB approximation to
the wave Eq. (14) yields the exact energy eigenvalues for the
solvable spherically symmetric potentials. In our approach, the
radial equation (17) has the centrifugal term $(l+\frac 12)^2\hbar^2/r^2$
for {\em any} spherically symmetric potential $V(r)$ because the
squared angular momentum eigenvalues $\vec M^2=(l+\frac 12)^2\hbar^2$
are obtained in a natural way from solution of the angular
semiclassical equation (18) with the use of the same WKB method. In
other words, we have shown that the Langer replacement $l(l+1)\ra
(l+\frac 12)^2$ requires the modification of the angular momentum.
This correction is universal for any spherically  symmetric potential
and no further corrections are necessary. \newpage

\noindent {\bf 5. Generalization and discussion} \medskip

The standard lowest-order WKB prescription reproduces the exact
energy levels for the one-dimensional harmonic oscillator and
three-dimensional harmonic oscillator in the Cartesian coordinates
$x,y,z$. But just these two problems are correctly formulated
in the framework of the semiclassical approach: in the Cartesian
coordinates $x,y,z$, the Schr\"odinger equation has the
required canonical form and coincides with the semiclassical one.

The required canonical form is (in the spherical coordinates) the
semiclassical Eq. (14), in which the centrifugal term has the form
$(l+\frac 12)^2\hbar^2/r^2$ for all spherically symmetric potentials. 
An analogous semiclassical wave equation can be written in the general 
case of the curvilinear coordinates $q_1(x,y,z)$, 
$q_2(x,y,z)$,$q_3(x,y,z)$:

\be \left[ \sum_{k=1}^3\left( \frac{-i\hbar }{g_{kk}}\frac \pa {\pa
q_k}\right) ^2\right] \tilde \psi (\vec q)=2m\left[ E-V(r)\right]
\tilde \psi (\vec q), \ee
where $g_{kk}$ are the elements of the metric tensor. The correlation
of the function $\tilde \psi (\vec q)$ with the wave function$\psi
(\vec q)$ of the Schr\"odinger equation is given by the formula (16).
The quasiclassical condition (11) implies that the momentum $\pa
S_0/\pa r$ is large enough, i.e., the quantum number $n$ takes large
values. At the same time, the WKB method yields exact eigenvalues for
{\em all} values of $n$. To disentangle this contradiction let us
return to Eq. (9) and show that the condition (11) can be
generalized.

In Eq. (9), the expression in the square brackets has a sense of
squared momentum. In order for the operator $[(-i\hbar)\pa /\pa
r]^2$ to be Hermitian this expression should be real. This is
possible if $\pa ^2S_0/\pa r^2\simeq 0$. What does this
condition mean? This implies the adiabatically slow alteration of the
derivative $\pa S_0/\pa r$, i.e.,

\be \frac{\pa S_0}{\pa r}\simeq const. \ee
Unlike the condition (11), the constraint (40) supplies the
hermiticity of the operator (12) and does not imply that the momentum
$\pa S_0/\pa r$ takes large values. Further, this constraint
anticipates the final result, i.e., discrete constant eigenvalues
$k_n$ of the operator (12).  Integrating (40), we obtain, for the
action $S_0(r)$, $S_0(r)=p_nr+$ const, where $p_n$ is the momentum
expressed via the energy eigenvalue $E_n$, $p_n= \sqrt{2m\left|E_n\right|
}$, and the final solution can be written in elementary functions. In
the region of the classical motion, where $p(r)>0$, this solution has
the form of a standing wave

\be \tilde R_n(r)=A\,\cos\left(\frac{p_nr}{\hbar} -\chi_1 -
\frac\pi 4\right), 
\ee
where $\chi_1$ is the value of the phase integral (24) at the turning
point $r_1$. Analogous solutions can be written for other one-dimensional
equations obtained after separation of Eq. (14). These solutions are
in agreement with the asymptotic solutions of the corresponding exact
solutions of the Schr\"odinger equation (1). \newpage

\noindent {\bf 6. Conclusion } \medskip

In conclusion, let us summarize the results obtained . Our approach
to the problem under consideration is different from known ones
\ci{Rep},\ci{Dunh}-\ci{Bru} in which the one-dimensional or radial
problems have been considered. In this work, we have considered an
approach concerning the application of the WKB method to the
three-dimensional problems in quantum mechanics. Whereas previous
workers were considering modification of the WKB method or were using
some transformations of the one-dimensional equations obtained after
separation, we start with the original three-dimensional problem. We
have shown that the main problem is to reduce the Schr\"odinger
equation to the correct canonical form, i.e., to an equation without
first derivatives.

The main result of this work is the derivation of the semiclassical
wave Eq.  (14) [or in general form (39)] appropriate in the
quasiclassical region; to do this, the basic principles of quantum
mechanics were used: the correspondence principle and the adiabatic
one. Unlike the Schr\"odinger equation (1), the semiclassical one
(14) results in another integral of motion, i.e., the squared angular
momentum $\vec M^2=(l+\frac 12)^2\hbar ^2$ . This means that the
centrifugal term in the radial equation (17) has the form
$(l+\frac 12)^2\hbar^2/r^2$ for {\em any} spherically symmetric potential
$V(r)$. It is important to emphasize that the squared angular
momentum eigenvalues $\vec M^2=(l+\frac 12)^2\hbar^2$ have been
obtained in our approach in a natural way from the solution of the
angular semiclassical equation (18) in the framework of the same WKB
method.  In other words, we have obtained the justification of the
Langer correction as the correction to the squared angular momentum
eigenvalues.

We have shown that the solution of the obtained wave equation (14) by 
the standard WKB method (to leading order in $\hbar$) gives the exact 
eigenvalues for {\em all} solvable spherically symmetric potentials.  
The corresponding eigenfunctions have the same behaviour as the 
asymptotes of the exact solutions. A generalization of the 
semiclassical equation for the arbitrary curvilinear coordinates 
$q_1(x,y,z)$, $q_2(x,y,z)$,$q_3(x,y,z)$ has been obtained.

We have considered here the three-dimensional problem in spherical
coordinates. To deal with the one-dimensional or other
multi-dimensional problems, one must, first of all, write the
equation under consideration in the correct canonical form.  For this
one should start from the corresponding classical Hamilton-Jacobi
equation and, using the correspondence principle (15), write a wave
equation. Then each of the one-dimensional equations obtained after
separation is solved by the WKB method.

Thus the standard leading-order WKB approximation is the appropriate
method to solve the semiclassical wave equation (14) obtained . We
have shown that quantization, the apparent form of the operators, and
many results of quantum mechanics (exact eigenvalues, correct
asymptotic behaviour of the semiclassical wave functions at the
origin and at infinity, and the correct phases of the WKB solutions) can
be obtained within the framework of the standard semiclassical
approach.

{\em Acknowledgment}. This work was supported in part by the
Belarusian Fund for Fundamental Researches.

\end{document}